%% file: arxiv-text.tex
\documentclass{article}
\usepackage{arxiv}

\usepackage[utf8]{inputenc} 
\usepackage[T1]{fontenc}    
\usepackage{hyperref}       
\usepackage{url}            
\usepackage{booktabs}       
\usepackage{amsfonts}       
\usepackage{nicefrac}       
\usepackage{microtype}      

\usepackage{algorithm}
\usepackage{algpseudocode}
\usepackage{amsmath}
\usepackage{amssymb}

\usepackage{graphicx}

\newenvironment{authorsummary}
{
  \centerline
  {\large \bfseries \scshape Author Summary}
  \begin{quote}
}
{
  \end{quote}
}

\title{Natural representation of composite data with replicated autoencoders}
\author{Matteo Negri\textsuperscript{1,2} \And Davide Bergamini\textsuperscript{2} \And Carlo Baldassi\textsuperscript{2} \And Riccardo Zecchina\textsuperscript{2} \And Christoph Feinauer\textsuperscript{2*} \\ \And
}
\date{\vspace{-0.5cm}
    \begin{flushleft} 
        \small 
        \textbf{1} Dept. Applied Science and Technology, Politecnico di Torino, Corso Duca degli Abruzzi 24, I-10129 Torino, Italy \\
        \textbf{2} Artificial Intelligence Lab, Institute for Data Science and Analytics, Bocconi University, via Sarfatti 25, I-20136 Milan, Italy \\
        \textbf{*} Corresponding Author: christoph.feinauer@unibocconi.it
    \end{flushleft}
}
\begin{document}
\maketitle

\begin{abstract}
Generative processes in biology and other fields often produce data that can be
regarded as resulting from a composition of basic features. Here we present
an unsupervised method based on autoencoders for inferring these basic
features of data. The main novelty in our approach is that the training is based on the optimization of the
`local entropy' rather than the standard loss, resulting in a more robust
inference, and enhancing the performance on this type of data considerably.
Algorithmically, this is realized by training an interacting system of
replicated autoencoders. We apply this method to synthetic and protein
sequence data, and show that it is able to infer a hidden representation
that correlates well with the underlying generative process, without
requiring any prior knowledge.
\end{abstract}

\begin{authorsummary}
Extracting compositional features from noisy data and identifying the corresponding generative models is a fundamental challenge across sciences.
The composition of elementary features can have highly non-linear effects which makes them very hard to identify from experimental data.

In biology, for instance, one challenge is to identify the key steps or components of molecular and cellular processes. Representative examples are the modeling of protein sequences as the composition of patterns influenced by phylogeny or the identification of gene clusters in which the presence of specific genes depends on the evolutionary history of the cell.

Here we present an unsupervised machine learning technique for the analysis of compositional data which is based on entropic neural autoencoders. Our approach aims at finding deep autoencoders that are highly invariant with respect to perturbations in the inputs and in the parameters. The procedure is efficient to implement and we have validated it both on synthetic and protein sequence data, where it can be shown that the latent variables of the autoencoders are non trivially correlated with the true underlying generative processes. Our results suggests  that the local entropy approach represents a general valuable tool for the extraction of compositional features in hard unsupervised learning problems.

\end{authorsummary}

\section{Introduction}
There are many examples of data that can be thought of as a composition of
basic features. For such data, an efficient description can often be constructed by a
-- possibly weighted -- enumeration of the basic features that are present in a
single observation.

As a first example, we could describe genomes of single organisms as a
composition of genes and gene clusters, where the presence or absence of
specific genes is determined by the evolutionary history and further reflected
in the presence or absence of functions and biochemical pathways the organism
has at its disposal \cite{mazzolini2018statistics, albalat2016evolution}.
Depending on the level of description, such a composition is not necessarily a
linear superposition of the basic features. It has recently been estimated, for
example, that due to horizontal gene transfer the genome of Homo Sapiens
outside of Africa is composed of $1.5\%-2.1\%$ of Neanderthal DNA
\cite{prufer2014complete} but no single genomic locus is actually a
superposition. Nonetheless, such a description conveys a lot of information: a machine learning algorithm that could be trained in an unsupervised manner on
a large number of genomes and automatically output such coefficients would be very valuable
in the field of comparative genomics \cite{hardison2003comparative}.

As a further example we can take the gene expression signature of a single
cell, which is determined by the activity of modules of genes that are
activated depending on cell identity and the environmental
conditions \cite{segal2003module}. Since there are far fewer such gene modules
than genes, the activity of these modules can be used as an efficient
description of the state of the cell. The inference of such modules based on
single cell genomic data and downstream tasks like clustering cells into
subtypes is an ongoing field of research \cite{trapnell2015defining}.

On a even more fine-grained level, there have been recently several successful
efforts to model protein sequence data as a composition of features that arise
from structural and functional constraints and are also influenced by phylogeny
\cite{tubiana2017emergence, tubiana2019learning, tubiana2019learningb}. This leads to several
possible patterns of amino acids for making up functional groups, or contacts
between amino acids, and the presence or absence of these patterns can be used
as features and inferred from aligned sequence data of homologous proteins.

There are also many examples of composite data outside of biology. An
immediate example are images that contain multiple objects.  The efficient
extraction of such objects, which can be seen as basic features, has important
applications, for example for self-driving cars \cite{redmon2016you}. In such
applications, one is of course also interested in the number and locations of
the objects, but a basic description of an image, using an enumeration of
objects present, can be part of a general pipeline.

As a final example, we note that in natural language processing documents are
often modeled as a mixture of topics, each of which gives a contribution to
different aspects of the document: for example, the authors of ref.~\cite{blei2003latent} use the
distribution of words. As in the case of genomes, the actual document is far
from being a superposition of the topics, but such a description is nonetheless
useful in fields like text classification.

A natural candidate model for finding efficient representations are
undercomplete, sparse autoencoders \cite{lecun2015deep}. These are multi-layer
neural networks that are trained (in an unsupervised fashion, i.e.~on unlabeled data)
to realize the identity function. Their goal is to learn a compressed parametrization
of the data distribution: to this end, the training is performed under the constraint
that the internal representation in a specific layer, called \textit{bottleneck}, is
sparse and low-dimensional. Under the assumption that only a few basic features
contribute to any given observation and that the number of basic features is smaller
than the dimension of the bottleneck, such an internal representation could be expected
to identify the basic features that describe the data.

In this work, we present evidence that it is indeed possible to find
representations of composite data in terms of basic features, but that this
process is very sensitive to both overfitting and underfitting: If the imposed
sparsity is not strong enough, the resulting representation does not correspond
to the basic features. If it is too strong, the dictionary of basic features is
not represented completely.

Therefore we present a modified version of the autoencoder, the
\textit{replicated autoencoder}, which is designed to find good solutions in
cases where overfitting is a danger. We test this hypothesis on synthetic and
on real, biological data. In both cases we find more natural representations of
the data using the replicated autoencoder.

\section{Results}

\subsection{Natural representations of composite data}

By composite data we mean data in which a single observation $\vec{x}$, represented by a vector of real numbers, can be roughly seen as a function $f$ of $D$ basic features $\vec{v}_d$ and real weights $\alpha_d$ for $d=1 \ldots D$:

\begin{align*}
\vec{x} = f(\vec{v}_1,\alpha_1,\ldots,\vec{v}_D,\alpha_D)
\end{align*}

The basic features $\vec{v}$ could be either one-hot encodings for categorical features or real valued vectors. We call the set of all $\vec{v}$ the \textit{dictionary} and $D$ the size of the dictionary. The weight $\alpha_d$ determines how strongly the basic feature $\vec{v}_d$ contributes to the observation $\vec{x}$ and could be either binary, encoding presence or absence of a basic feature, or a positive real number that quantifies the contribution. The easiest version of composite data would be a linear superposition, but we do not limit ourselves to this case. In fact, the synthetic data shown below is not a linear superposition of the basic features. We also do not assume that the equality holds \textit{exactly} but allow for noise and other factors to influence the observation $\vec{x}$.

In this work we study multi-layer neural networks trained in an unsupervised fashion on composite data. In such a network, a natural way of representing data that is a composition of basic features is to use the activations of one of the hidden layers as a representation of the input and match each of the hidden units of this layer with one basic feature. For a specific input coming from this data, only the neurons corresponding to the basic features present in that input should then show a significant activation, and the activation should be correlated to the corresponding weight. Under the assumption that the number of features included in each example is much lower than the total number of possible features, we expect such a representation of composite data to be sparse. We do not assume that the size of the dictionary $D$ or the basic features $\vec{v}_d$ are known, but infer them from the data.

\subsection{Model}

\begin{figure}
\includegraphics[width=\linewidth]{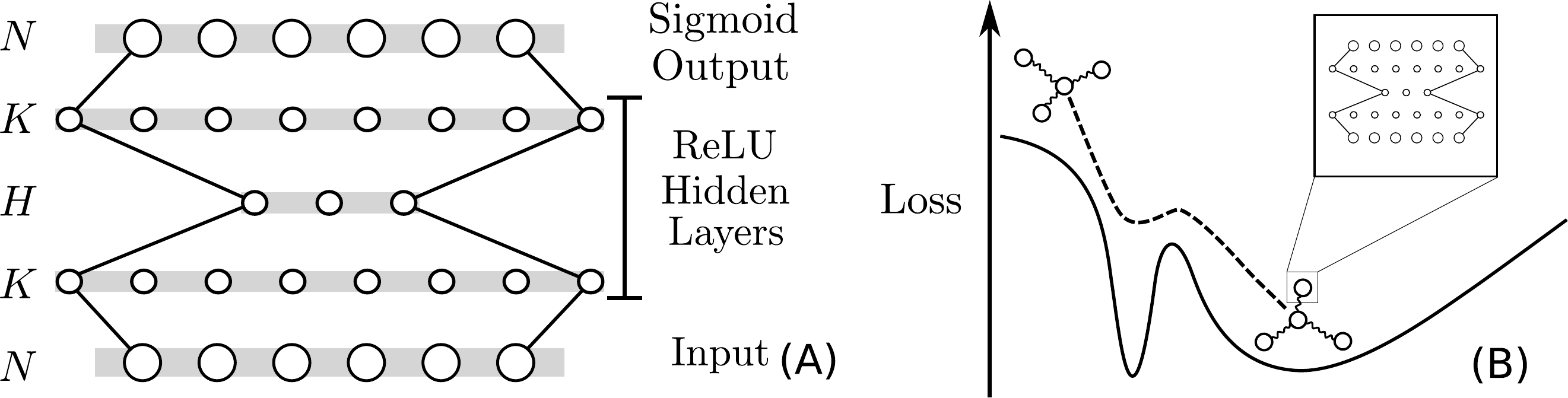}
\caption{{\bf Basic schemes of our model.} {\bf (A)} Scheme of the basic autoencoder architecture used throughout the paper. {\bf (B)} Sketch of the robust optimization procedure. It depicts the evolution of the system of interacting replicas on the loss landscape. In the sketch the system has three replicas interacting with a center (each circle represents an autoencoder). The system as a whole avoids narrow minima and ends up in a high-local-entropy region (a wide minimum).}
\label{fig:model}
\end{figure}

We train feed-forward autoencoders (AE) with stochastic gradient descent (SGD), minimizing the reconstruction error $L_{\mathrm{err}}$.
The basic AE model we consider is made of an input layer, three hidden layers and an output layer,  made respectively by $N$, $K$, $H$, $K$ and $N$ units, with $K>N>H$, see fig.~\ref{fig:model}. The smallest layer with size $H$ is the bottleneck. We use the activations in the bottleneck as the representation of the input.

In each layer except the last one we use as a activation function the rectified linear unit (ReLU), defined as $\mathrm{ReLU}(x):=\max(0,x)$. In order to obtain a sparse representation in the bottleneck, we add an L1 penalty for the activations $h_l$ of the neurons of the central hidden layer: $L_{\mathrm{reg}}=\frac{1}{N} \sum_{K=1}^H |h_K|$.

To summarize, for a single autoencoder (S-AE) we use the loss function
\begin{equation}
L_{\mathrm{tot}}=L_{\mathrm{err}}+\gamma L_{\mathrm{reg}},
\label{eq:regularizer}
\end{equation}
where $\gamma$ is the regularizer strength. Higher values of  $\gamma$ enforce lower activations of the units in the hidden layer and higher overall sparsity (for a detailed discussion on the general effects of this regularizer, see for example ref.~\cite{goodfellow2016deep}).

We observe that with higher $\gamma$ there are more units that show little to no activation on any input pattern in the training set: The auto-encoder shuts down some units in a trade-off between the two regularization terms in the loss function. We call the number of active units $D^*$ the \textit{inferred dictionary size}, and $H-D^*$ is the number of deactivated units. We infer $\gamma$ and therefore $D^*$ from the data (see below).

\subsection{Replicated systems and overfitting}

If the data is indeed composite in nature, we expect a representation that captures the true underlying contribution of features to generalize well. Therefore we take special measures to avoid overfitting, which would decrease generalization performance and would constitute evidence that the current representation in the hidden units is not corresponding to the basic features in the data. We note that in neural networks, overfitting has been connected to sharp minima in the loss function \cite{baldassi2016unreasonable, chaudhari2016entropy, baldassi2019shaping}. To avoid such minima, we modify the optimization procedure to prefer weights that are in the vicinity of other weights that have low loss, which is a measure of the flatness of the loss landscape. Borrowing physics terminology, we call the number of low loss weights around a specific set of weights the "local entropy" of the weights. The analogy is that while in physics the entropy is a measure of how many states a system in equilibrium can occupy when the states are weighted with their probabilities, in our case the local entropy determines how many sets of weights are around some specific solution when weighted with the likelihood on the training set. Intuitively, one expects weights with a high local entropy to generalize well since they are more robust with respect to perturbations of the parameters and the data and therefore less likely to be an artifact of overfitting. In fact, such flat minima have already been found to have better generalization properties for some deep architectures \cite{chaudhari2016entropy}. We call the optimization procedure that finds such minima \textit{robust optimization}.

More precisely, the local (free) entropy of a certain configuration of the weights $\vec{w}^*$ is defined as \cite{baldassi2016unreasonable}:
\begin{equation}
\Phi\left(\vec{w}^*; \beta, \lambda\right) = \log \sum_{\vec{w}} \exp\left(-\beta \left(L_{\mathrm{tot}}\left(\vec{w}\right) + \lambda d(\vec{w},\vec{w}^*)^2\right)\right),
\end{equation}
where the function $d$ measures the distance between the weights: several choices are possible, but in the rest of the work we use exclusively the euclidean distance. The parameter $\lambda$ controls indirectly the locality, i.e.~the size of the portion of landscape around $\vec{w}^*$ that we are considering (a larger $\lambda$ corresponds to a smaller radius). The parameter $\beta$ has the role of an inverse temperature in physics, and it controls indirectly the amount of flatness required of the local landscape (a larger $\beta$ corresponds to flatter landscapes).

Computing the local entropy is expensive and impractical in most cases. However, as described in detail in ref.~\cite{baldassi2016unreasonable}, if we use the negative local entropy $-\Phi$ as an energy function (i.e. as the objective function that we wish to optimize) with an associated fictitious "inverse temperature" $R$ that we choose to be a positive integer, the canonical partition function of the system is amenable to an equivalent description that can be implemented in a straightforward way: We add $R$ replicas of our model, $\left(\vec{w}^{\left(r\right)}\right)_{r=1}^R$, and we add an interaction between each replica $r$ and the central (original) configuration $\vec{w}^*$ that forces them to be at a certain distance. We thus end up with the new replicated objective function
\begin{equation}
\label{eq:L_R}
L_{\mathrm{R}} = \sum_{r=1}^R L_{\mathrm{tot}}^{(r)} +  \lambda\sum_{r=1}^R d(\vec{w}^{(r)},\vec{w}^*)^2,
\end{equation}
where $L_{\mathrm{tot}}^{(r)}$ is the total loss of the replica $r$. It is important at this stage to observe that the canonical physical description presupposes a noisy optimization process where the amount of noise is regulated by some inverse temperature $\beta$, while in this work (following ref.~\cite{baldassi2016unreasonable}) we will be relying on the noise provided by SGD instead, thereby using the mini-batch size and the learning rate as "equivalent" control parameters. Relatedly, we should also note that, although the interaction term is purely attractive, the replicas won't collapse unless the coupling coefficient $\lambda$ is very large, due to the presence of noise in the optimization. Thus, in our protocol, the coefficient $\lambda$ is initialized to some small value and gradually increased at each training epoch.

Besides the analytical argument, the intuitive reason why this procedure achieves more robust optimization results is that the interaction will prevent replicas to remain trapped in bad minima of the loss: if one of the replicas finds an overfitted solution and this overfitting is associated with a sharp minimum, it is likely that the other replicas will not be at the same minimum, but at a higher point in the loss function. The overfitted replica will be pulled out of the sharp minimum as the interaction term grows (see figure \ref{fig:model} for a sketch).

The robust optimization protocol that we have used throughout this work can be then summarized as follows (additional details can be found in the Materials and Methods section \nameref{sec:algo}). We train $R$ autoencoders with different initializations coupled with a central autoencoder $\vec{w}^*$, which we call the center. Every replica is trained on batches from the training set with normal SGD, but we add a gradually increasing coupling term between every replica and the central autoencoder, see Eq.~(\ref{eq:L_R}). At the end of the training procedure, we have $R$ trained replicas and one center. All of these $R+1$ models are autoencoders that can be used for prediction or representation. We typically discard all replicas and only use the center. We call an autoencoder that is trained using this procedure a \emph{replicated} autoencoder (R-AE). In the rest of this work, we ask if this robust optimization is helpful for finding a natural representations of composite data. We test this idea first on synthetic data where we control the generative process and then extend the approach to protein sequence data. In the latter case the exact generative process is unknown, but a coarse approximation to the basic features can be found in the taxonomic labels.

\subsection{Synthetic data}
Following ref.~\cite{mezard2017mean}, we generate synthetic datasets $X=\{\vec{x}^{\mu}\}_{\mu=1}^M$ of examples $\vec{x}^{\mu}$ obtained as superpositions of basic features, modeled as follows. We consider a dictionary of basic features $\{\vec{v}_d\}_{d=1}^D$, where $D$ is the size of the dictionary. In this setup, we choose $\vec{v}_d$ as a random binary ($0$ or $1$) sparse vector of length $N$. We use binary weights $\alpha_d^{\mu}$ to control the contribution of the basic feature $\vec{v}_d$ on the observation $\vec{x}^{\mu}$ and set only a small number of the weights to $1$ for each observation. The final observation is defined to be
\begin{equation}
	\vec{x}^{\mu}=\min \left( 1, \sum_{d=1}^D \alpha^{\mu}_d \vec{v}_{d} \right) \ \ \ \ \mu \in \left\{1\dots M\right\}
\label{eq:combd}
\end{equation}

Note that this is not a simple linear superposition due to the element-wise $\min$ function. The purpose of this way of generating data is to let all basic features have a potential impact on every observation while keeping the task of inferring their contributions and the basic features themselves non-trivial.
A possible representation of the data is one where each feature $\vec{v}_d$ in the dictionary corresponds to a single hidden unit in the central layer of the autoencoder. We call this the \emph{natural} representation of the synthetic dataset. This representation needs $D$ hidden units. For this reason we expect the autoencoder to be able to find the natural representation when $H \ge D$, given that an appropriate value for $\gamma$ has been used.
Additional details on synthetic data generation can be found in the Materials and Methods section \nameref{sec:synth}.

\subsection{R-AE versus S-AE on synthetic data}

We train a single autoencoder (S-AE) and a replicated autoencoder (R-AE) on the synthetic data. We compare the reconstruction performance on unseen examples, the regularization loss and the ability to infer the basic features based on the hidden units of the bottleneck.

\begin{figure}
\includegraphics[width=\linewidth]{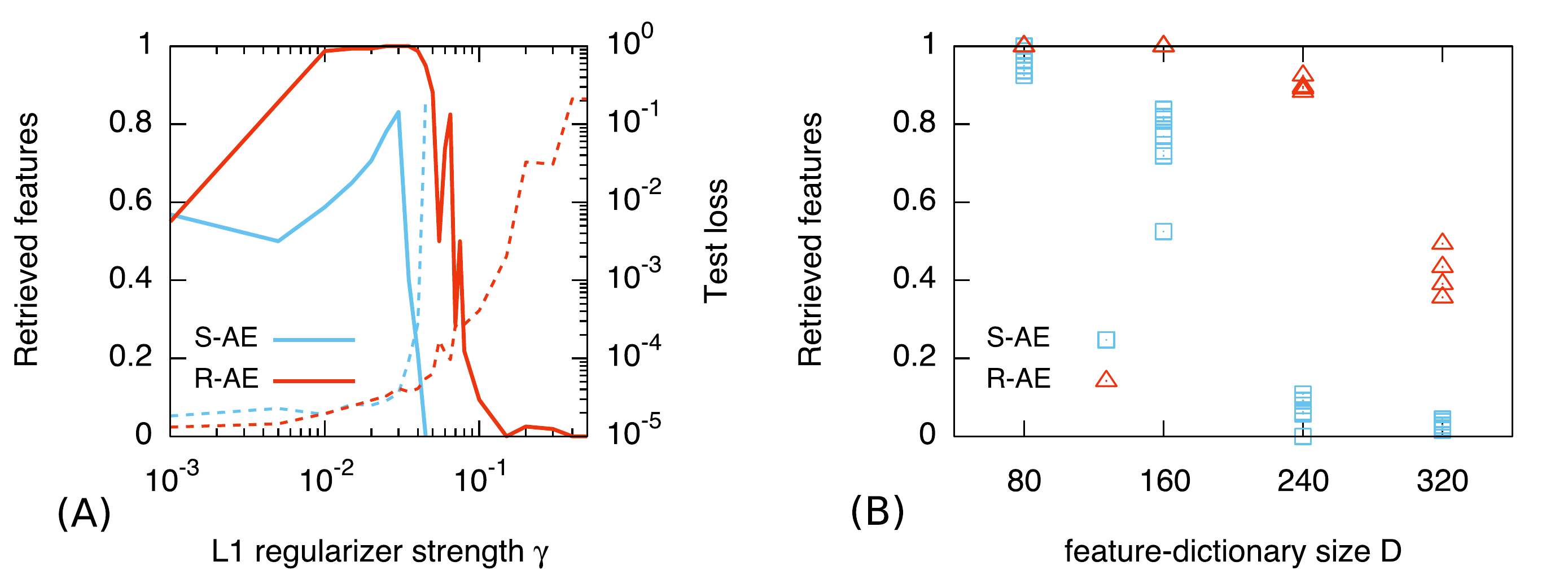}
	\caption{{\bf The AE is able to retrieve all the features of a sufficiently small dictionary.} \textbf{(A)} The test loss  (\textbf{dashed line}, right y-axis) increases slowly with $\gamma$, up to a certain knee point $\gamma^*$, which corresponds to the point where the fraction of retrieved features has a maximum (\textbf{solid line}, left y-axis); for $\gamma > \gamma^*$ the performance quickly deteriorates. The curves are obtained with one execution of the training for each value of $\gamma$. The dimension of the dictionary is fixed to $D=160$.
\textbf{(B)} The performance of the AE trained on a dataset $X_D$ (y-axis) depends on the dimension of the dictionary $D$ (x-axis): the retrieval of features is better for smaller dictionaries, and the robust AE is able to fully retrieve bigger dictionaries (for $D=80$  both single and robust AE retrieve 100\% of the features, while for $D=160$ only the robust AE is able to do so). For each $D$ the plot shows 9 results with S-AE and 4 with R-AE, each one corresponding to different realizations of the same training procedure. The regularizer strength is set in the proximity of the knee point, namely $\gamma=3\cdot10^{-2}$.}
\label{fig:synth_result}
\end{figure}

The striking difference between the R-AE and the S-AE is that the R-AE is able to achieve a better reconstruction performance at high sparsity, in the region where $\gamma\gtrsim 0.03$ (see fig.~\ref{fig:synth_result}A). This is connected to the observation that the R-AE has a number of active units $D^*$ that is significantly larger than the S-AE while keeping a similar L1 norm for most inputs. This might sound paradoxical, but we recall here that $D^*$ is the number of units in the bottleneck that show a significant activation for at least one input from the training set. This is not directly suppressed by the L1 regularization on the bottleneck, which penalizes cases in which many units are activated for a single input. There are thus different ways to realize the same overall L1 norm. The S-AE deactivates more units completely, while using a larger fraction of the remaining active units on the inputs on average. The R-AE, on the other hand, deactivates fewer units completely, but using a smaller fraction of the active units for every input. Another way of stating this is that the R-AE uses representations that are more distributed over all available units and keeps $D^*$ closer to $D$ (see \nameref{S-fig:rank_plot_comparison} for an example of this behaviour).

Both S-AE and R-AE are able to retrieve all the features of a sufficiently small dictionary, if $\gamma$ is chosen such that $D^*\simeq D$. At the same time we observe that for the R-AE, the range of $\gamma$ for which this is true is significantly wider than for the S-AE (fig. \ref{fig:synth_result}A, solid line).
If we plot the loss as a function of $\gamma$, we observe that it grows slowly up to a certain knee point $\gamma^*$ (fig.~\ref{fig:synth_result}A, dashed line). This point coincides with the maximum number of retrieved features. This can be interpreted as a phase transition between overfitting and underfitting, and for $\gamma > \gamma^*$ the performance deteriorates quickly.

In general, the retrieval of features is better for smaller dictionaries for both models, but for larger dictionary sizes the R-AE retrieves a higher number of features, see fig.~\ref{fig:synth_result}B: for $D=80$  both the S-AE and the R-AE retrieve 100\% of the features, while for $D=160$ only the R-AE is able to do so. For $D\ge240$ the R-AE finds $\sim40\%$ more features than the S-AE.

\subsection{Biological data: protein families}

In this section we test the capability of the R-AE to infer basic features on real data. We use sequence data of homologous proteins because they allow a reasonable interpretation of composition: due to co-evolution of residues that are part of structural contacts or functional groups, certain patterns of amino acids arise. These patterns can be exploited for the prediction of contacts with the structure of a single protein \cite{morcos2011direct, cocco2018inverse}, infer protein interaction networks \cite{cong2019protein, feinauer2016inter} and paralogs \cite{gueudre2016simultaneous, bitbol2016inferring}, model evolutionary landscapes \cite{figliuzzi2015coevolutionary} and predict pathogenicity of mutations in humans \cite{hopf2017mutation, feinauer2017context}. Since these patterns are inheritable, we expect their presence to be partly determined by the phylogenetic history of the organism and therefore to be correlated with its taxonomy. We therefore argue that a `natural' representation of an amino acid sequence should be correlated with taxonomy of the organism.

We thus proceed as follows. We consider a wide variety of protein families (see the Materials and Methods section \nameref{sec:protein}), and we use aligned sequences in a one-hot encoding as the input of the autoencoders. Each family is partitioned in train set, test set and validation set in the proportion 80\%-10\%-10\%. We then test two different measures of correlation between the representations of the S-AE and the R-AE of the sequences with their taxonomic labels. Note that, analogously to the case of synthetic data, the training of the autoencoders is agnostic about these labels.

The behavior of the autoencoders trained on protein sequence data is qualitatively similar to what we saw for synthetic data: there is always a knee point in the curve of the loss (both train and test) as a function of $\gamma$, see \nameref{S-fig:all_peaks_00}, \nameref{S-fig:all_peaks_01}, \nameref{S-fig:all_peaks_02}. We expect that the range of values around the knee point corresponds to a representation that is close to the underlying biology.

We determine the knee point $\gamma^*$ for a given  protein family by fitting the error curve (not directly the loss) on the test set by two connected line segments and then use the point where they intersect as $\gamma^*$. All the subsequent analysis is done on the validation set, using the autoencoder with the identified $\gamma^*$. See Materials and Methods, Section \nameref{subsec:knee} for a more detailed description of the procedure.

We measure the taxonomic information captured by the hidden layer in two ways: First, in analogy with synthetic data, under the very hopeful hypothesis that each taxonomic label corresponds to a single hidden unit. We test this idea in the next paragraph. Secondly, we ask how well a clustering of the sequences based on the hidden representations correlates with the taxonomic labels in comparison to a clustering based directly on the amino acid sequences.

Since the taxonomic classification is modeled by a tree, we consider the labels
as organized by their depth $d$, that is their distance from the root of the
tree. For example, the root has $d=0$, the label 'Bacteria' has $d=1$,
'Proteobacteria' has $d=2$. Every label is associated with one or more
sequences in the training set and every sequence corresponds to several labels.
The labels near the root are the most populated, while the labels deeper in the
tree are sparsely populated.  We expect the labels in the first few levels to
be more correlated with the hidden unit since deeper labels correspond to only
a few sequences in the training set.  For these reasons we restricted the
following analysis to labels up to depth $d=5$, with the additional condition
that they must contain at least 20 sequences from the training set.

\subsubsection{Neuron-Taxon correlation}

Given a taxonomic label indexed by $l$, we consider the binary variable $y_{l}(s^{\mu})$ that, for each sequence $s^{\mu}$ in the MSA, is equal to $1$ if the sequence belongs to that taxon and is equal to $0$ otherwise; then, after the AE has been trained, we compute (on the training set) the correlation matrix $C_{l,k}$ between the variables $\{y_{l}\}$ and the activations $\{h_{k}\}$ of the hidden units. For every label $l$ we select the most correlated unit $k^*(l)$. Then we define a score $Q$ as the average correlation (on the test set) of the most correlated units for every label:
\begin{equation}
Q:=\frac{1}{L}\sum_{l=1}^L C_{l,k^*(l)}
\label{eq:Q}
\end{equation}
where $L$ is the total number of labels considered in a MSA.

\begin{figure}
\includegraphics[width=\linewidth]{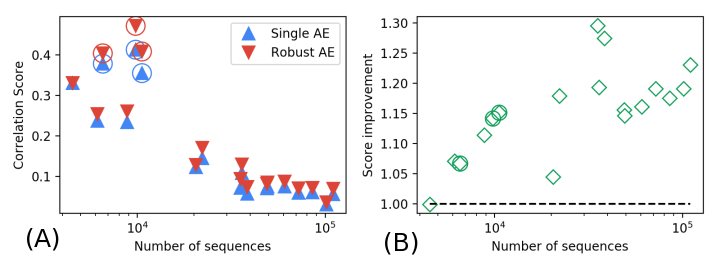}
	\caption{{\bf The robust AE consistently captures more biological information in most of the protein families considered.} \textbf{(A)}  The panel shows an aggregate score of correlation between the hidden units of the network and the taxonomic labels present in each protein family (eq.~(\ref{eq:Q})); the families are show on the x-axis according to their number of sequences. The circled points correspond to ribosomal domains, which appear to be the datasets with the highest performance of our method. \textbf{(B)}  The panel shows, for each family, the score improvement gained by training the robust AE respect to training the single AE.}
\label{fig:bio_result_1}
\end{figure}

The results are shown in Fig.~\ref{fig:bio_result_1}: R-AE consistently finds a higher score than S-AE. It is useful to note the general trend of this score: the more sequences in the dataset, the worse the score. Additionally, the protein families of ribosomal domains have a much higher score, which is probably due to the fact that ribosomes are well sampled (see the \nameref{sec:discussion} section for more on this).

\subsubsection{Clustering data in the latent space}

For a given label $l$ at depth $d$, we consider the sub-labels $l'$ at depth $d+1$ branching from $l$; we select the subset of the training set corresponding to the label $l$, then we compute the centroids of the clusters corresponding to the sub-labels $l'$ by averaging the sequences with that sub-label. Given a new sequence from the test set belonging to $l$, we assign the sub-label $l'$ according to the closest centroid.
In order to perform this clustering procedure on disjoint subsets in such a way that the accuracies are independent from each other, we fix the depth $d$ and consider only labels $l$ found at that depth. We choose $d=2$, because it provides the most variety of sub-labels with a high number of examples in the protein families we considered.

First we run this procedure using the original sequences, the same ones on which we trained the AEs. Then we repeat the clustering using, for each sequence, its representation in terms of the hidden units of the AEs. We ask whether this representation improves the accuracy of the clustering, depending on whether we use the representation from R-AE or S-AE.

\begin{figure}
\centering
\includegraphics[width=0.8\linewidth]{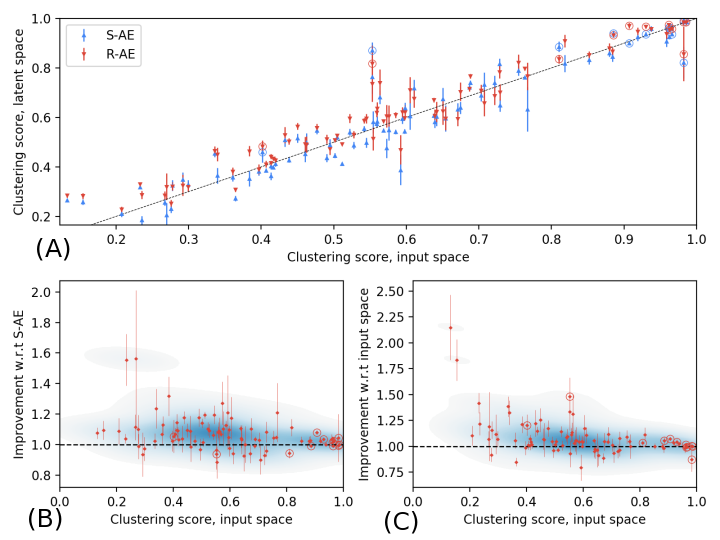}
\caption{{\bf The representation of the AE improves a clustering algorithm.} \textbf{(A)} The panel shows the accuracy of a clustering algorithms run into the latent space of the neural network versus the accuracy of a clustering algorithms run on the original data point. Each point corresponds to a label in a protein family: given the subset of a protein family corresponding to that label, we consider the task of clustering that subset according to the subcategories of that label. \textbf{(B,C)} The panels show two 2D density plot of the score impovement as a function of the score on the original data points, respect to S-AE (B) and to the input space (C).  }
\label{fig:bio_result_2}
\end{figure}

The results are shown in fig.~\ref{fig:bio_result_2}: the representation learned by R-AE does improve the accuracy for the majority of labels both respect to S-AE (bottom-left panel) the to input space (bottom-right panel).

\section{Discussion\label{sec:discussion}}

In this work we have presented a method to extract representations of composite data that connects to the structure of the underlying generative process. To this end, we combined two techniques that allowed us to recover such representations from the bottleneck of autoencoders trained on the composite data: The first is a regularization that forces the autoencoder to use a sparse representation. The second is the replicating of the autoencoder, which changes the properties of the solutions found. We showcased the method on two different datasets. In the first dataset, where we controlled the generative process, we showed that the replication allows to extract the underlying basic features also in cases where the sparsity constraints are too strong for a single autoencoder. After a closer analysis, we found that replication enables the system to effectively disentangle basic features and specialize parts of the internal representations.

In a second step, we applied the same method to protein sequence data. Since patterns of amino acids are inheritable, we used the correlation between the extracted representations and the phylogenetic labels as a metric for assessing the quality of the representation. We found that the replication of the autoencoder resulted in representations that are closer to biological reality and that the qualitative characteristics of the loss function and internal representations are similar to autoencoders trained on synthetic data.

One intriguing observation is that the point where the internal representation becomes correlated with the basic features is identifiable: The knee point in the loss curve in dependence of the regularization parameter corresponds to the peak performance in feature retrieval. For synthetic data we were able to verify this directly. Near this knee point, each hidden unit represented one basic feature. This also allowed us to infer the number of basic features present in the data (i.e.~its inherent dimensionality). For protein sequence data, we observed that the internal representation becomes correlated with taxonomic labels at the knee point. After this knee point, the loss deteriorates quickly, indicating that the autoencoder starts dropping important information from the internal representation.

Interestingly, we found that feature retrieval on synthetic data became more difficult for increasing dictionary sizes. This could be addressed either by using a larger and therefore more expressive architecture or by using a larger training set. We generally expect the size of the training set necessary for the extraction of the basic features to scale with the size of the dictionary \cite{davenport2016overview,arora2015simple}.

Regarding the difficulty of the feature extraction task, we found a similar behavior on the protein families: Families with more sequences also contain a higher number of labels and are expected to have a wider variety of features. It is further noteworthy that the correlation between taxonomic labels and internal representations was more pronounced for ribosomal domains than for other families with a similar number of sequences (see fig.~\ref{fig:bio_result_1}, circled markers). This is probably due to the fact that ribosomes are well studied systems in many species and that in the databases we used there are more species per sequence for these domains (see tab.~\ref{tab:PF}). This indicates that a well balanced data-set is an additional factor in the inference of basic features from composite data.

\section{Conclusion}

In conclusion, we have shown that replicated autoencoders are capable of finding representations of composite data that are meaningful in terms of the underlying biological system. We believe the approach to be very general, since we used no prior knowledge about the biology involved. The work we presented here encourages us to believe that the method could be useful for other data in biology where the representations and basic features extracted might lead to new biological insights. As one example for a possible direction of future research we point to the increasing number of measurements coming from single-cell transcriptomics \cite{zeisel2015cell}. In these data, the basic features are conceptually clearer than in protein sequence data and we suspect that representations of cell states in terms of gene expression modules would be found. Such representations could in turn be used to cluster cell types or analyze pathologies like Alzheimer's Disease \cite{mathys2019single}.

\section{Materials and Methods}

\subsection{Synthetic data}
\label{sec:synth}

We generate synthetic data points according to eq.~(\ref{eq:combd}) with the following characteristics: each example has $784$ components, we generate training sets with $M_{\mathrm{train}}=60000$ examples and a test set with $M_{\mathrm{test}}=10000$ examples. We used four different datasets with dictionary sizes $D=80$, $160$, $240$ and $320$.

The architecture (fig.~\ref{fig:model}) is fixed: we used intermediate hidden layers with $K=1000$ and a bottleneck with $H=400$ for all our experiments with synthetic data.

We chose the feature vectors $\vec{v}_d$ to be random with binary ($0$ or $1$) independent entries, with a fixed average fraction $p_v$ of non-zero components. The coefficients $\alpha^{\mu}_d$ are also binary, sparse and random: they were generated with a probability $p_d$ of being non-zero. However, in order to make the retrieval problem sufficiently difficult, for each pattern $\vec{x}^{\mu}$ we ensured that it contained \textit{at least} three features (i.e.~we discarded and resampled those that didn't meet the criterion $\sum_d \alpha^{\mu}_d \ge 3$).
The generation of a dataset is therefore parametrized by $N$, $M_{\mathrm{train}}$, $M_{\mathrm{test}}$, $D$, $p_v$ and $p_d$. In this work, we fixed the sparsity of the features at $p_v=0.1$ and the sparsity of the coefficients at $p_d=0.01$. We always chose $M\gg D$.

Since we work with binary patterns, the activation function of the output layer is chosen to be $\mathrm{Sigm}(x):=1/(1+e^{-x})$, which sets the range of each output unit between $0$ and $1$.
The loss function of choice for these datasets is mean square error (MSE), which is simply the squared difference between a unit in the input layer and the corresponding unit in the output layer, summed over all the units.

\subsection{Protein data}
\label{sec:protein}

We considered 18 protein families from the PFAM database (tab.~\ref{tab:PF}) selected according a number of criteria: we want many types of proteins represented, as well as families covering many different partitions of the tree of life; additionally, we chose families with a sufficient number of sequences and species, varying the ratio between these two numbers.

\begin{table}[h]
\begin{tabular}{lccc|lccc}
DATASET    & n. seq  & n. species & n. amm & DATASET    & n. seq & n. species & n. amm \\
\hline
PF01978.19					& 4531   & 1806      & 68  & PF04545.16 & 35976  & 8384      & 50  \\
PF09278.11 				& 6117   & 2867      & 65  & PF00805.22 & 38453  & 3485      & 40  \\
\textbf{PF00444.18}    	& 6551   & 5971      & 38  & PF07676.12 & 48848  & 6060      & 38  \\
PF03459.17 				& 8823   & 4066      & 64  & PF00356.21 & 49284  & 5450      & 46  \\
\textbf{PF00831.23}    	& 9782   & 9209      & 57  & PF03989.13 & 60674  & 8153      & 48  \\
\textbf{PF00253.21}    	& 10577 & 8650      & 54  & PF01381.22 & 72011  & 9760      & 55  \\
PF03793.19 				& 20495 & 4026      & 63  & PF00196.19 & 85219  & 6666      & 57  \\
PF10531.9  					& 22080 & 7683      & 58  & PF00353.19 & 101177 & 2304      & 36  \\
PF02954.19 				& 35339 & 5079      & 42  & PF04542.14 & 110168 & 8385      & 71  \\
\end{tabular}

\medskip

Source: \url{https://pfam.xfam.org/}
\caption{\label{tab:PF}List of dataset used for training the AE, listed by their number of sequences. The protein families of ribosomal domains are highlighted; notice that, for these families, the ratio of the number of different species over the number of sequences is higher than for the other families.}
\end{table}

Given a sequence  $S=\{a_i\}_{i=0}^{A}$ of length $A$ to the AE, we represent each amino-acid $a_i$ with a 21-components one-hot encoding: each input sequence is thus a binary vector of length $N=A\times 21$, and the entire dataset with a matrix $(M, N)$. The architecture is rescaled according to the sequence length $A$: we set $K=1.1\times N$, and the number of units in the bottleneck to $H=0.4\times N$.

Since each amino acid is a categorical variable represented by a one-hot encoding, a common way to compute the reconstruction error $L^{(i)}$ for a single amino acid $a_i$ is the cross entropy between the input and the output. To do this, we consider the 21 units $\vec{z}^{(i)}$ in the output layer that describe the site $i$, then we apply a softmax operation so that each unit can be interpreted as a probability
\begin{equation}
\mathrm{Softmax(\vec{z})}:=\frac{e^{z_j}}{\sum_{j=1}^{21}e^{z_j}}
\end{equation}
and finally we compute the cross entropy
\begin{equation}
L^{(i)}=-z_{j^*}+\log \sum_{j=1}^{21}e^{z_j}
\end{equation}
where $j^*$ is the index corresponding to the true value of the amino acid. The complete loss function is the summation of the cross entropies for each amino acid of the sequence:
\begin{equation}
L=\sum_{i=1}^{A} L^{(i)}.
\label{eq:CELoss}
\end{equation}
Here we choose a linear activation function for the units in the output layer.

\subsection{Learning algorithm}
\label{sec:algo}

The algorithm we use to train R-AE consists in iterating two alternating steps: a step of SGD on each replica computed on its own reconstruction loss, followed by a step in which each replica is pushed towards the center and the center towards the replicas. In practice this procedure is similar to elastic-averaging SGD \cite{zhang2015deep}, which in turn is related to the optimization of local entropy \cite{baldassi2016unreasonable}. The pseudo-code for the algorithm is sketched in alg.~\ref{alg:replicatedSGD}.

\begin{algorithm}[]
\caption{Training procedure for replicated autoencoder}
\label{alg:replicatedSGD}
\hspace*{\algorithmicindent} \textbf{Input:} current weights $\vec{w}^{(r)},\vec{w}^*$\\
\hspace*{\algorithmicindent} \textbf{Hyper-parameters:} batch size, learning rate $\eta$, coupling $\lambda$
\begin{algorithmic}[1]
\For{$i = 1,\dots,\mathrm{steps}$}
	\For{$r = 1,\dots,\mathrm{replicas}$}
	\State $x \gets \mathrm{minibatch}[i,r]$
	\State $\vec{w}^{(r)} \gets \vec{w}^{(r)} - \eta \vec{\nabla}_w L(\vec{w}^{(r)}; x)$
	\EndFor
	\For{$r = 1,\dots,\mathrm{replicas}$}
	\State $\vec{w}^{(r)} \gets \vec{w}^{(r)} - \lambda (\vec{w}^{(r)}-\vec{w}^*)$
	\State $\vec{w}^* \gets \vec{w}^* + \lambda (\vec{w}^{(r)}-\vec{w}^*)$
	\EndFor
\EndFor
\end{algorithmic}
\end{algorithm}

We impose an exponential scheduling on the coupling $\lambda$ between the replicas and the center, namely we take $\lambda(t)=\lambda_0(1+\lambda_1)^t$, where $t$ is the time step of the training in units of epochs.

The training of S-AE is performed with the same procedure with just one replica and setting $\lambda=0$.

In order to set the values for the many hyperparameters of these algorithms, we decided to select one prototype case among the synthetic data and one among the proteins data, and then to proceed by trial and error in order to find a regime in which the training converges and has good performance; once we found these values, we assumed that the general performances should not be sensitive to the fine-tuning of the hyperparameters: for this reason we used the same set of hyperparameters for every synthetic dataset and the other set of hyperparameters for all the protein families. We observed them to work well in the majority of cases.

For synthetic data we set $\eta=2.5\cdot10^{-4},\, \lambda_0=4\cdot10^{-2},\, \lambda_1=3\cdot10^{-2}$ and trained for $600$ epochs. For all protein data data we set $\eta=5,\, \cdot10^{-4},\, \lambda_0=8\cdot10^{-3},\, \lambda_1=3\cdot10^{-2}$ and trained for $300$ epochs. The training epochs sufficient for reaching convergence with respect to the training loss. The batch size was fixed to 50 for all trainings.


We did not use any momentum, which resulted in a deterioriation of performances
across every region of parameters and non-convergence.  The reason for this
behavior could be that the loss landscape for this optimization problem appears
to be highly non-convex, especially when the regularization approaches the
region near $\gamma^*$; momentum-related techniques, on the other hand, are
designed to work well when the loss landscape is sufficiently smooth
\cite{goodfellow2016deep}.

\subsection{Knee Point Identification}
\label{subsec:knee}
Part of our approach is identifying the knee point $\gamma^*$ in the loss curve. To this end, we consider the reconstruction error curve on the test set in dependence of $\gamma$. The curve has two parts, separated by the knee point: A slow increase in reconstruction performance (decrease in error) and a drastic decrease in reconstruction performance (drastic increase in error) when $\gamma$ becomes too high. We fit the the region around the knee point with the function:

\begin{equation}
f\left(\gamma\right)=\begin{cases}
a_{1}\left(\gamma-\gamma^*\right)+b & \textrm{if }\gamma<\gamma^{*}\\
a_{2}\left(\gamma-\gamma^*\right)+b & \textrm{if }\gamma\ge\gamma^{*}\\
\end{cases}
\label{eq:knee}
\end{equation}
which is simply the equation of two straight lines passing from the same point at $\gamma^*$. From the fit over the four parameters $a_1,\, a_2,\, b,\, \gamma^*$ we obtain the estimation of the position of the knee point.

We use the error curve (the number of wrong amino acids in the reconstruction) rather than the loss directly, since the error curve is better approximated by two line segments and therefore easier fitted by our approach, leading to better approximations of the knee point.

The knee point is different for each protein and we expected it to be also different for S-AE and R-AE. Empirically, however, we obtained the best results across all the protein families by using the $\gamma^*$ of the S-AE also for the R-AE.

\section{Supporting information}

\paragraph*{S1 Fig.}
\label{S1_Fig}
{\bf Example trajectories of the loss during the training.} Ten trajectories are shown for S-AE and three for R-AE. The panels on the left show the train loss, the right ones show the test loss. Here we show a case where S-AE and R-AE have the same performance (D=160, top line) and one case where R-AE has a much better performance (D=240, top line). Note that the improvement is greater for the test loss, showing that R-AE generalizes better. These trajectories refer to Fig. 2B in the main text. The regularizer strength is set in the proximity of the knee point, namely $\gamma=3\cdot10^{-2}$.

\paragraph*{S2 Fig.}
\label{S-fig:rank_plot_comparison}
{\bf Examples of average activation of the bottleneck neurons.} There are different ways to realize the same overall L1 norm of the units in the bottleneck layer. The figure shows rank plots for different AE. On the x-axis there are the hidden units ranked by their average activation: the units on the right are the most active on average and the ones on the far right are the ones that are always deactivated (their signal is next to zero across all the dataset). On the y-axis there is the average activation of the units. In the high sparsity case  we can see that S-AE deactivates more units completely, while R-AE, on the other hand, deactivates fewer units completely. This effect disappears at lower sparsity, far from the knee point of the loss curve. The dataset used for these result is PF01978.19.

\paragraph*{S3 Fig.}
\label{S-fig:all_peaks_00}
{\bf Loss and score curves for all the proteins considered (part 1 of 3).} The behavior of the autoencoders trained on protein sequence data is qualitatively similar to what we saw for synthetic data: there is always a knee point in the curve of the loss as a function of $\gamma$, corresponding to the maximum correlation with the taxonomic labels.

\paragraph*{S4 Fig.}
\label{S-fig:all_peaks_01}
{\bf Loss and score curves for all the proteins considered (part 2 of 3).}

\paragraph*{S5 Fig.}
\label{S-fig:all_peaks_02}
{\bf Loss and score curves for all the proteins considered (part 3 of 3).}

\section{Funding Acknowledgements}
CB and RZ acknowledge ONR Grant N00014-17-1-2569.

\bibliographystyle{unsrt}  
\bibliography{biblio}

\include{supp-content}

\end{document}

%% file: supp-content.tex
\graphicspath{{supplemental/}}
\setcounter{figure}{0}
\renewcommand{\thefigure}{S\arabic{figure}}
\section{Supplemental Information}
\subsection{Synthetic data}


\begin{figure}[h]
\centering
\includegraphics[scale=0.5]{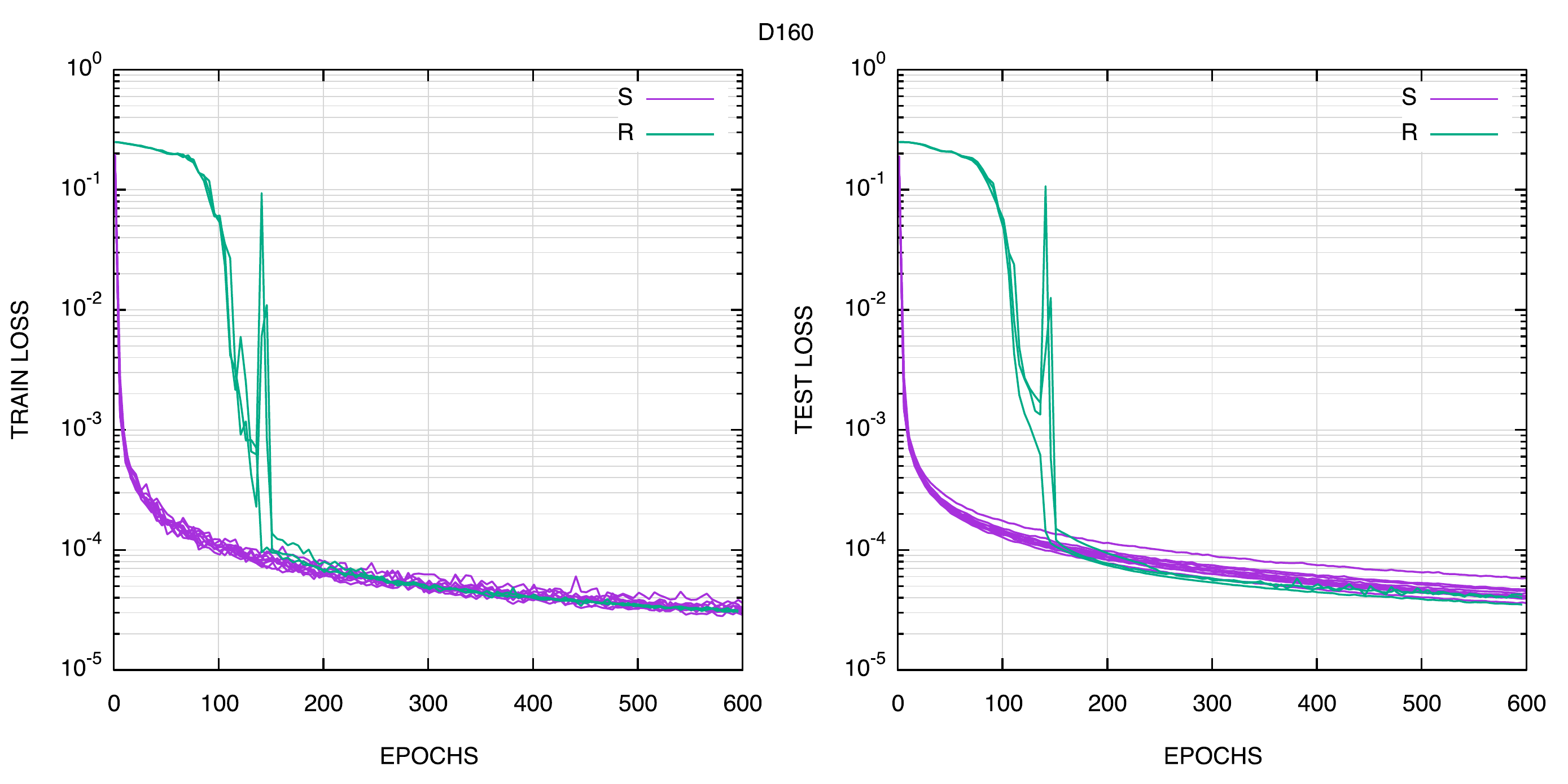}
\includegraphics[scale=0.5]{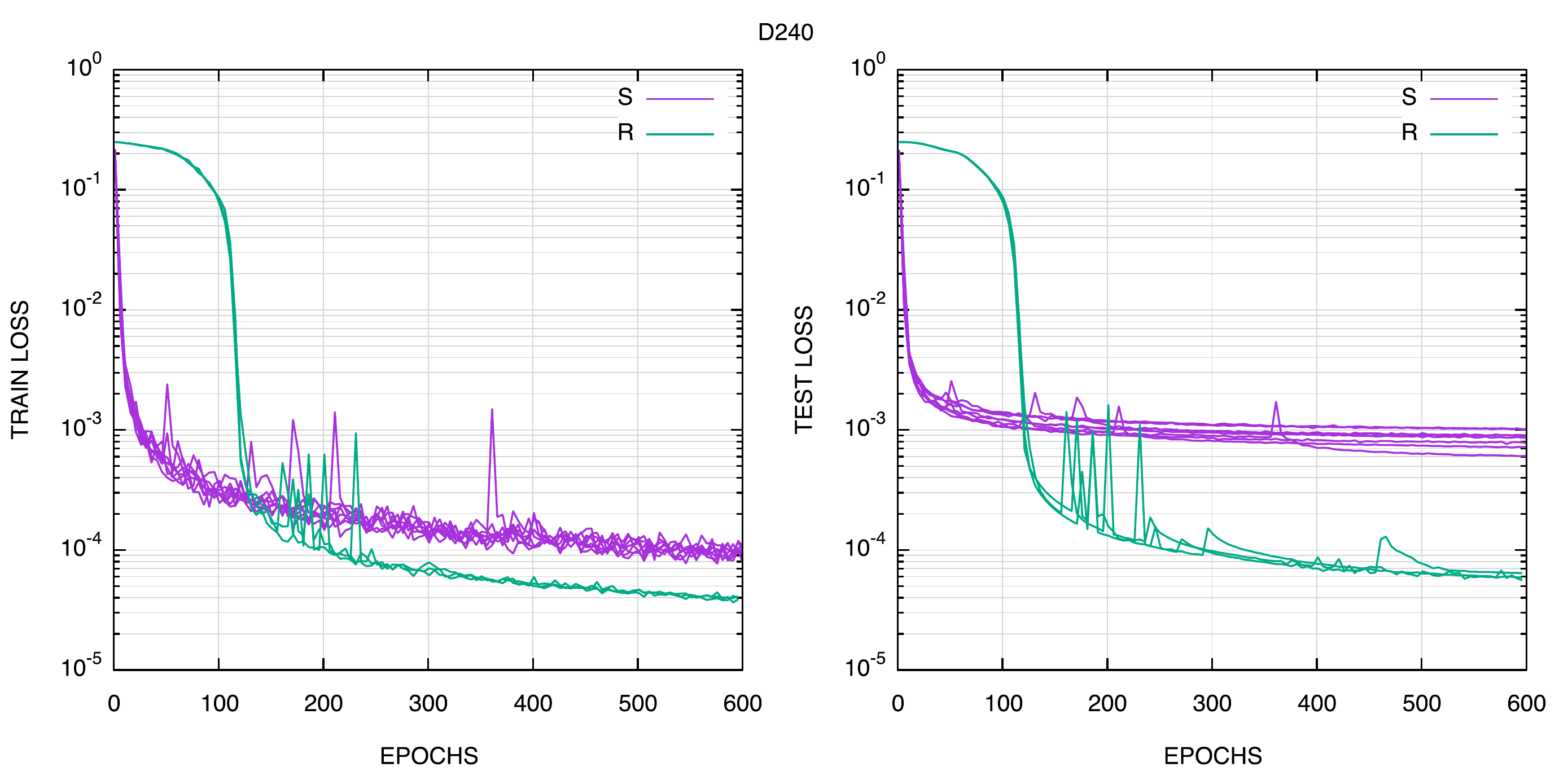}
\caption{ Example trajectories of the loss during the training; ten trajectories are shown for S-AE and three for R-AE. The panels on the left show the train loss, the right ones show the test loss. Here we show a case where S-AE and R-AE have the same performance (D=160, top line) and one case where R-AE has a much better performance (D=240, top line). Note that the improvement is greater for the test loss, showing that R-AE generalizes better. These trajectories refer to Fig. 2B in the main text. The regularizer strength is set in the proximity of the knee point, namely $\gamma=3\cdot10^{-2}$.}
\label{fig:dinamics_D240}
\end{figure}

\pagebreak
\subsection{Biological data: protein families}

\begin{figure}[h]
\centering
\includegraphics[scale=1.0]{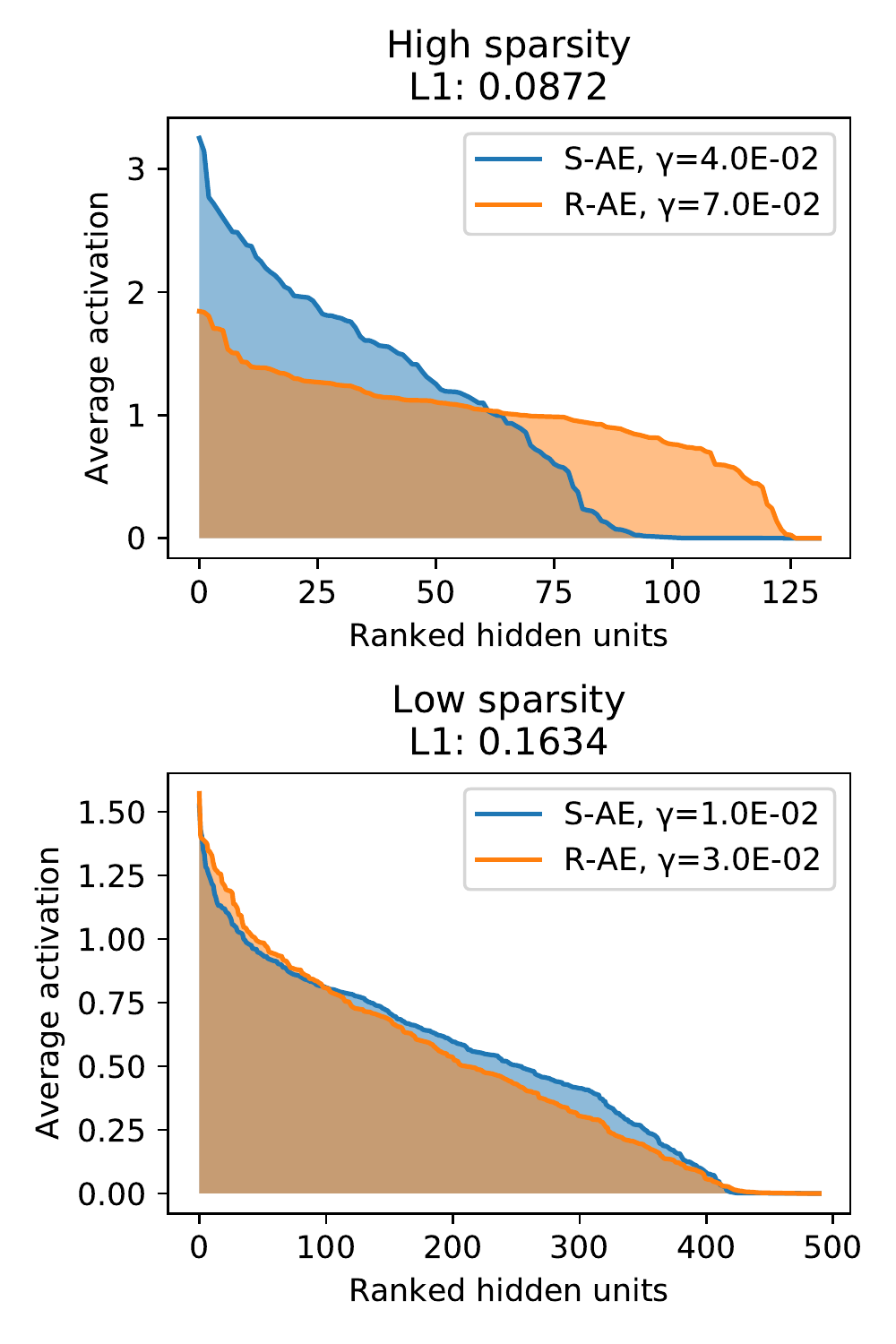}
\caption{There are different ways to realize the same overall L1 norm of the units in the bottleneck layer. The figure shows rank plots for different AE. On the x-axis there are the hidden units ranked by their average activation: the units on the right are the most active on average and the ones on the far right are the ones that are always deactivated (their signal is next to zero across all the dataset). On the y-axis there is the average activation of the units. In the high sparsity case  we can see that S-AE deactivates more units completely, while R-AE, on the other hand, deactivates fewer units completely. This effect disappears at lower sparsity, far from the knee point of the loss curve. The dataset used for these result is PF01978.19.}
\label{fig:rank_plot_comparison}
\end{figure}

\begin{figure}[h]
\centering
\includegraphics[scale=0.5]{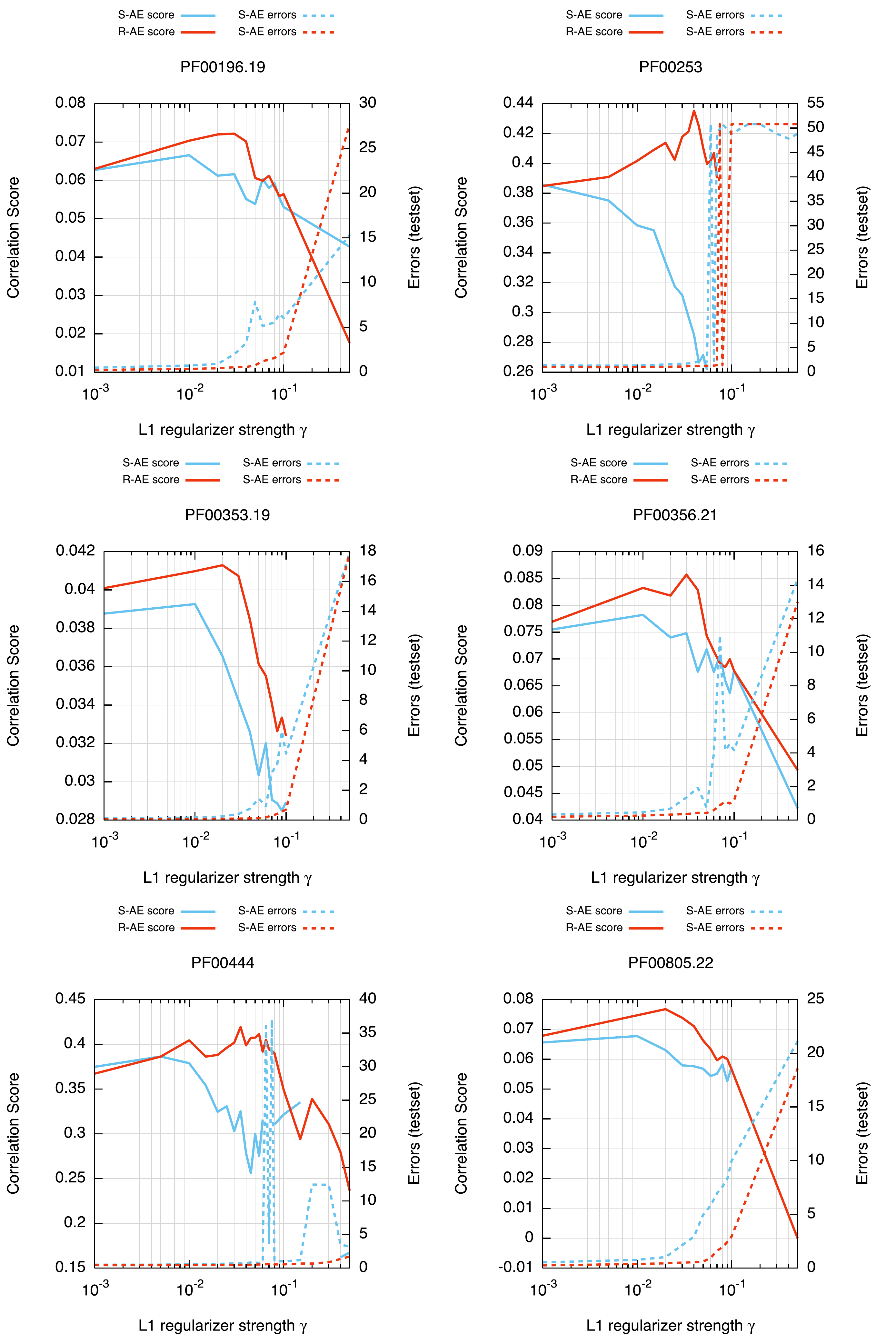}
\caption{Part 1:The behavior of the autoencoders trained on protein sequence data is qualitatively similar to what we saw for synthetic data: there is always a knee point in the curve of the loss as a function of $\gamma$, corresponding to the maximum correlation with the taxonomic labels.}
\label{fig:all_peaks_00}
\end{figure}

\begin{figure}[h]
\centering
\includegraphics[scale=0.5]{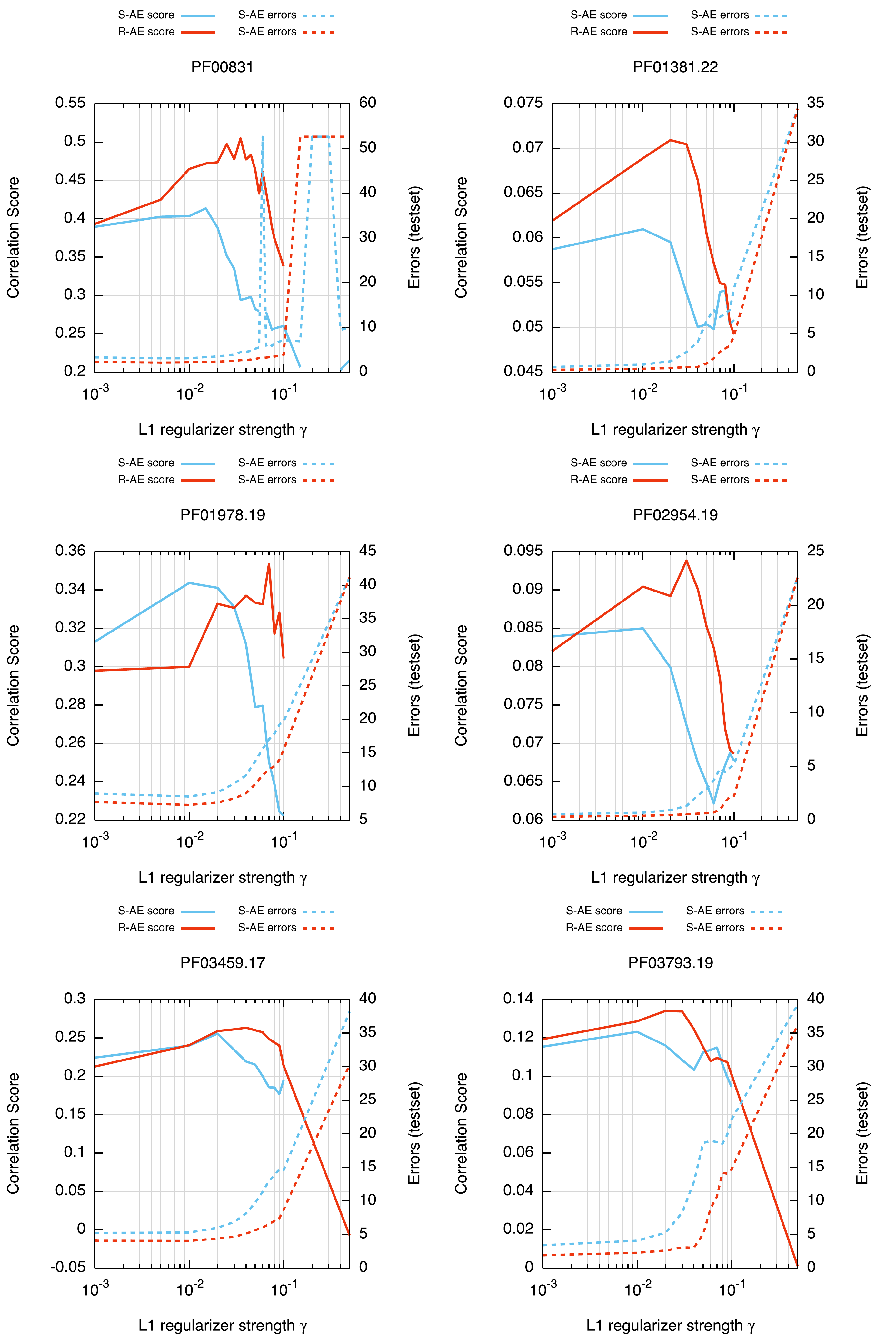}
\caption{Part 2}
\label{fig:all_peaks_01}
\end{figure}

\begin{figure}[h]
\centering
\includegraphics[scale=0.5]{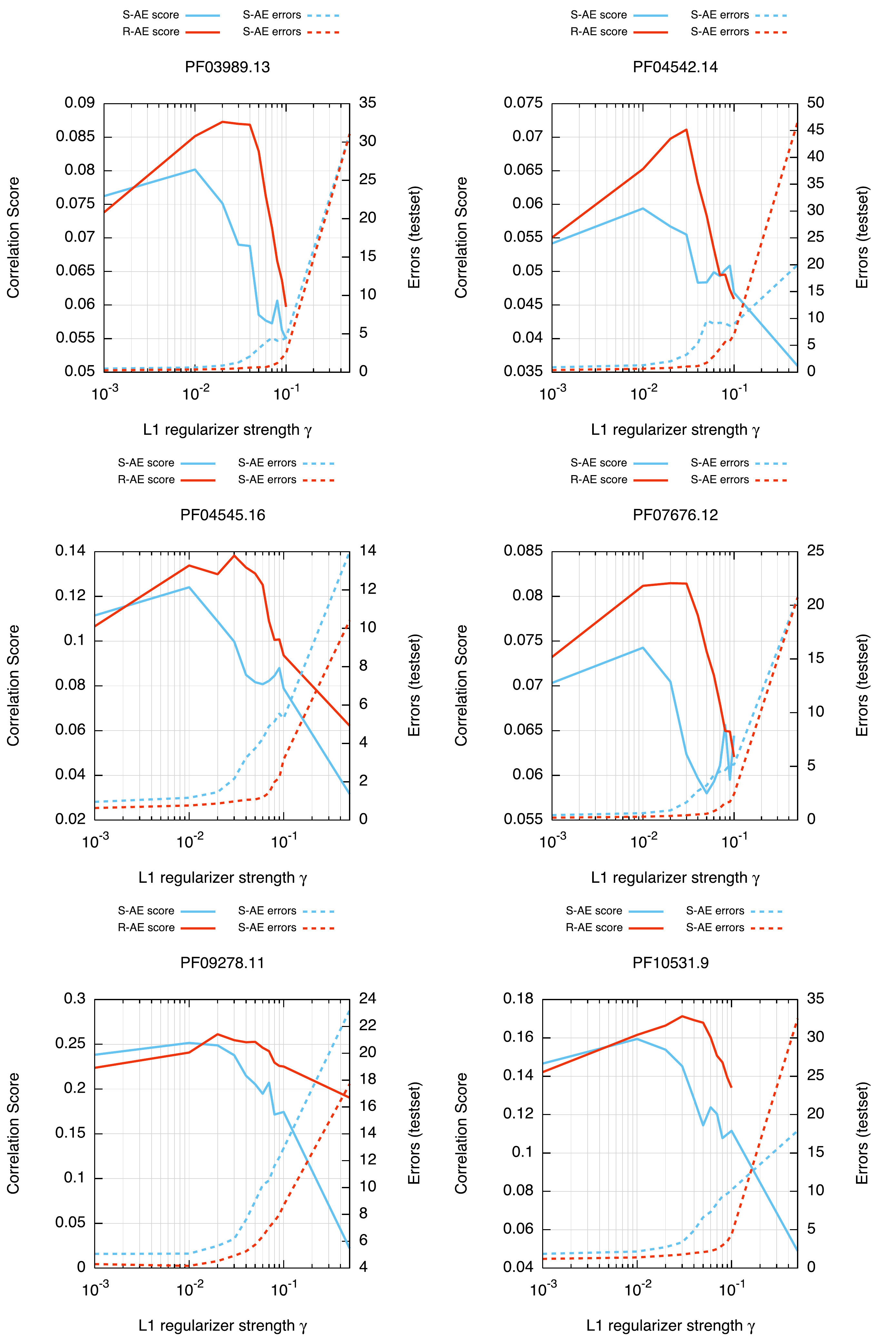}
\caption{Part 3}
\label{fig:all_peaks_02}
\end{figure}

%% file: arxiv-text.bbl
\begin{thebibliography}{10}

\bibitem{mazzolini2018statistics}
Andrea Mazzolini, Marco Gherardi, Michele Caselle, Marco~Cosentino Lagomarsino,
  and Matteo Osella.
\newblock Statistics of shared components in complex component systems.
\newblock {\em Physical Review X}, 8(2):021023, 2018.

\bibitem{albalat2016evolution}
Ricard Albalat and Cristian Ca{\~n}estro.
\newblock Evolution by gene loss.
\newblock {\em Nature Reviews Genetics}, 17(7):379, 2016.

\bibitem{prufer2014complete}
Kay Pr{\"u}fer, Fernando Racimo, Nick Patterson, Flora Jay, Sriram
  Sankararaman, Susanna Sawyer, Anja Heinze, Gabriel Renaud, Peter~H Sudmant,
  Cesare De~Filippo, et~al.
\newblock The complete genome sequence of a neanderthal from the altai
  mountains.
\newblock {\em Nature}, 505(7481):43, 2014.

\bibitem{hardison2003comparative}
Ross~C Hardison.
\newblock Comparative genomics.
\newblock {\em PLoS biology}, 1(2):e58, 2003.

\bibitem{segal2003module}
Eran Segal, Michael Shapira, Aviv Regev, Dana Pe'er, David Botstein, Daphne
  Koller, and Nir Friedman.
\newblock Module networks: identifying regulatory modules and their
  condition-specific regulators from gene expression data.
\newblock {\em Nature genetics}, 34(2):166, 2003.

\bibitem{trapnell2015defining}
Cole Trapnell.
\newblock Defining cell types and states with single-cell genomics.
\newblock {\em Genome research}, 25(10):1491--1498, 2015.

\bibitem{tubiana2017emergence}
J{\'e}r{\^o}me Tubiana and R{\'e}mi Monasson.
\newblock Emergence of compositional representations in restricted boltzmann
  machines.
\newblock {\em Physical review letters}, 118(13):138301, 2017.

\bibitem{tubiana2019learning}
J{\'e}r{\^o}me Tubiana, Simona Cocco, and R{\'e}mi Monasson.
\newblock Learning compositional representations of interacting systems with
  restricted boltzmann machines: Comparative study of lattice proteins.
\newblock {\em arXiv preprint arXiv:1902.06495}, 2019.

\bibitem{tubiana2019learningb}
J{\'e}r{\^o}me Tubiana, Simona Cocco, and R{\'e}mi Monasson.
\newblock Learning protein constitutive motifs from sequence data.
\newblock {\em eLife}, 8:e39397, 2019.

\bibitem{redmon2016you}
Joseph Redmon, Santosh Divvala, Ross Girshick, and Ali Farhadi.
\newblock You only look once: Unified, real-time object detection.
\newblock In {\em Proceedings of the IEEE conference on computer vision and
  pattern recognition}, pages 779--788, 2016.

\bibitem{blei2003latent}
David~M Blei, Andrew~Y Ng, and Michael~I Jordan.
\newblock Latent dirichlet allocation.
\newblock {\em Journal of machine Learning research}, 3(Jan):993--1022, 2003.

\bibitem{lecun2015deep}
Yann LeCun, Yoshua Bengio, and Geoffrey Hinton.
\newblock Deep learning.
\newblock {\em nature}, 521(7553):436, 2015.

\bibitem{goodfellow2016deep}
Ian Goodfellow, Yoshua Bengio, and Aaron Courville.
\newblock {\em Deep learning}.
\newblock MIT press, 2016.

\bibitem{baldassi2016unreasonable}
Carlo Baldassi, Christian Borgs, Jennifer~T Chayes, Alessandro Ingrosso, Carlo
  Lucibello, Luca Saglietti, and Riccardo Zecchina.
\newblock Unreasonable effectiveness of learning neural networks: From
  accessible states and robust ensembles to basic algorithmic schemes.
\newblock {\em Proceedings of the National Academy of Sciences},
  113(48):E7655--E7662, 2016.

\bibitem{chaudhari2016entropy}
Pratik Chaudhari, Anna Choromanska, Stefano Soatto, Yann LeCun, Carlo Baldassi,
  Christian Borgs, Jennifer Chayes, Levent Sagun, and Riccardo Zecchina.
\newblock Entropy-sgd: Biasing gradient descent into wide valleys.
\newblock {\em arXiv preprint arXiv:1611.01838}, 2016.

\bibitem{baldassi2019shaping}
Carlo Baldassi, Fabrizio Pittorino, and Riccardo Zecchina.
\newblock Shaping the learning landscape in neural networks around wide flat
  minima.
\newblock {\em arXiv preprint arXiv:1905.07833}, 2019.

\bibitem{mezard2017mean}
Marc M{\'e}zard.
\newblock Mean-field message-passing equations in the hopfield model and its
  generalizations.
\newblock {\em Physical Review E}, 95(2):022117, 2017.

\bibitem{morcos2011direct}
Faruck Morcos, Andrea Pagnani, Bryan Lunt, Arianna Bertolino, Debora~S Marks,
  Chris Sander, Riccardo Zecchina, Jos{\'e}~N Onuchic, Terence Hwa, and Martin
  Weigt.
\newblock Direct-coupling analysis of residue coevolution captures native
  contacts across many protein families.
\newblock {\em Proceedings of the National Academy of Sciences},
  108(49):E1293--E1301, 2011.

\bibitem{cocco2018inverse}
Simona Cocco, Christoph Feinauer, Matteo Figliuzzi, R{\'e}mi Monasson, and
  Martin Weigt.
\newblock Inverse statistical physics of protein sequences: a key issues
  review.
\newblock {\em Reports on Progress in Physics}, 81(3):032601, 2018.

\bibitem{cong2019protein}
Qian Cong, Ivan Anishchenko, Sergey Ovchinnikov, and David Baker.
\newblock Protein interaction networks revealed by proteome coevolution.
\newblock {\em Science}, 365(6449):185--189, 2019.

\bibitem{feinauer2016inter}
Christoph Feinauer, Hendrik Szurmant, Martin Weigt, and Andrea Pagnani.
\newblock Inter-protein sequence co-evolution predicts known physical
  interactions in bacterial ribosomes and the trp operon.
\newblock {\em PloS one}, 11(2):e0149166, 2016.

\bibitem{gueudre2016simultaneous}
Thomas Gueudr{\'e}, Carlo Baldassi, Marco Zamparo, Martin Weigt, and Andrea
  Pagnani.
\newblock Simultaneous identification of specifically interacting paralogs and
  interprotein contacts by direct coupling analysis.
\newblock {\em Proceedings of the National Academy of Sciences},
  113(43):12186--12191, 2016.

\bibitem{bitbol2016inferring}
Anne-Florence Bitbol, Robert~S Dwyer, Lucy~J Colwell, and Ned~S Wingreen.
\newblock Inferring interaction partners from protein sequences.
\newblock {\em Proceedings of the National Academy of Sciences},
  113(43):12180--12185, 2016.

\bibitem{figliuzzi2015coevolutionary}
Matteo Figliuzzi, Herv{\'e} Jacquier, Alexander Schug, Oliver Tenaillon, and
  Martin Weigt.
\newblock Coevolutionary landscape inference and the context-dependence of
  mutations in beta-lactamase tem-1.
\newblock {\em Molecular biology and evolution}, 33(1):268--280, 2015.

\bibitem{hopf2017mutation}
Thomas~A Hopf, John~B Ingraham, Frank~J Poelwijk, Charlotta~PI Sch{\"a}rfe,
  Michael Springer, Chris Sander, and Debora~S Marks.
\newblock Mutation effects predicted from sequence co-variation.
\newblock {\em Nature biotechnology}, 35(2):128, 2017.

\bibitem{feinauer2017context}
Christoph Feinauer and Martin Weigt.
\newblock Context-aware prediction of pathogenicity of missense mutations
  involved in human disease.
\newblock {\em arXiv preprint arXiv:1701.07246}, 2017.

\bibitem{davenport2016overview}
Mark~A Davenport and Justin Romberg.
\newblock An overview of low-rank matrix recovery from incomplete observations.
\newblock {\em IEEE Journal of Selected Topics in Signal Processing},
  10(4):608--622, 2016.

\bibitem{arora2015simple}
Sanjeev Arora, Rong Ge, Tengyu Ma, and Ankur Moitra.
\newblock Simple, efficient, and neural algorithms for sparse coding.
\newblock {\em CoRR}, abs/1503.00778, 2015.

\bibitem{zeisel2015cell}
Amit Zeisel, Ana~B Mu{\~n}oz-Manchado, Simone Codeluppi, Peter L{\"o}nnerberg,
  Gioele La~Manno, Anna Jur{\'e}us, Sueli Marques, Hermany Munguba, Liqun He,
  Christer Betsholtz, et~al.
\newblock Cell types in the mouse cortex and hippocampus revealed by
  single-cell rna-seq.
\newblock {\em Science}, 347(6226):1138--1142, 2015.

\bibitem{mathys2019single}
Hansruedi Mathys, Jose Davila-Velderrain, Zhuyu Peng, Fan Gao, Shahin
  Mohammadi, Jennie~Z Young, Madhvi Menon, Liang He, Fatema Abdurrob, Xueqiao
  Jiang, et~al.
\newblock Single-cell transcriptomic analysis of alzheimer’s disease.
\newblock {\em Nature}, page~1, 2019.

\bibitem{zhang2015deep}
Sixin Zhang, Anna~E Choromanska, and Yann LeCun.
\newblock Deep learning with elastic averaging sgd.
\newblock In {\em Advances in Neural Information Processing Systems}, pages
  685--693, 2015.

\end{thebibliography}
